\documentclass[cernpreprint,atlasdraft=false,texlive=2016,USenglish,texmf]{atlasdoc}
 
\usepackage{atlaspackage}
\usepackage{atlasbiblatex}
 
\usepackage{xcolor}
\definecolor{green2}{RGB}{2,137,0}
\usepackage{float}
\usepackage{rotating}
\usepackage{multirow}
\usepackage[11pt]{moresize}
\usepackage{etoolbox}
\usepackage{feynmp-auto}
\usepackage{adjustbox}
\usepackage{hyperref}
\usepackage{subfiles}
 
\usepackage{atlasphysics}
\usepackage{atlashepprocess}
\usepackage{atlasheavyion}
 
\graphicspath{{logos/}{figures/}}
 
\usepackage{ANA-HION-2019-05-PAPER-defs}

\AtlasTitle{Medium-induced modification of $Z$-tagged charged particle yields in Pb+Pb collisions at $5.02$~$\text{Te\kern -0.1em V}$\ with the ATLAS detector}

\AtlasAbstract{
The yield of charged particles opposite to a $Z$ boson with large transverse momentum ($p_\mathrm{T}$) is measured in $260$~pb$^{-1}$ of $pp$ and $1.7$~nb$^{-1}$ of Pb+Pb collision data at $5.02$~$\text{Te\kern -0.1em V}$\ per nucleon pair recorded with the ATLAS detector at the Large Hadron Collider. The $Z$ boson tag is used to select hard-scattered partons with specific kinematics, and to observe how their showers are modified as they propagate through the quark--gluon plasma created in Pb+Pb collisions. Compared with $pp$ collisions, charged-particle yields in Pb+Pb collisions show significant modifications as a function of charged-particle $p_\mathrm{T}$ in a way that depends on event centrality and $Z$ boson $p_\mathrm{T}$. The data are compared with a variety of theoretical calculations and provide new information about the medium-induced energy loss of partons in a $p_\mathrm{T}$ regime difficult to measure through other channels.}

\AtlasRefCode{HION-2019-05}
 
\PreprintIdNumber{CERN-EP-2020-148}

\arXivId{2008.09811}

\AtlasJournalRef{Phys. Rev. Lett. 126, 072301}
\AtlasDOI{10.1103/PhysRevLett.126.072301}

\hypersetup{pdftitle={ATLAS document},pdfauthor={The ATLAS Collaboration}}

\begin{document}
 
\maketitle

Collisions of heavy nuclei at ultrarelativistic energies at the Large Hadron Collider (LHC) and the Relativistic Heavy Ion Collider (RHIC) are understood to produce an extended region of hot and dense matter where partons exist in a deconfined state known as the quark--gluon plasma (QGP).
The high density of unscreened color charges in the QGP causes the showers of hard-scattered partons with large transverse momentum (\pt) to be modified as they traverse the medium~\cite{Busza:2018rrf}.
These modifications are observed in measurements of dijet and photon--jet momentum imbalance~\cite{HION-2012-11,Chatrchyan:2012nia,HION-2018-06,Sirunyan:2017qhf}, and in jet fragmentation functions~\mbox{\cite{HION-2017-04,CMS-HIN-12-013}}.
 
The large integrated luminosity of Pb+Pb collisions delivered during LHC Run 2 has enabled measurements of jets produced in association with a high-\pt \Zboson boson. At leading order, the $Z$ boson and the jet are produced back-to-back in the azimuthal plane, with equal \pt.
Since \Zboson bosons and their decay leptons, or similarly, photons, do not participate in the strong interaction and are not modified by the QGP~\cite{HION-2012-07,HION-2018-10}, they provide an estimate of the \pt and azimuthal direction of the partner hard-scattered parton before the developing shower is modified through interactions with the QGP~\cite{Kartvelishvili:1995fr,Neufeld:2012df}.
Measurements of photon-tagged fragmentation functions at the LHC~\cite{HION-2017-08,CMS-HIN-16-014} and photon--hadron correlations at RHIC~\cite{Adare:2012qi,STAR:2016jdz} used this feature to perform detailed studies of jet quenching.
At fixed $p_\mathrm{T}$, jets balancing $Z$ bosons and photons arise from processes with different Q$^2$, and can test the sensitivity of the energy loss process to parton virtuality. Additionally, the use of isolated photons at low photon-\pt ($\lesssim$ 60~\GeV) is difficult due to the large hadron-decay background, motivating the use of \Zboson bosons.
A measurement of $Z$+jet production with $\iptz > \SI{60}{\GeV}$ by CMS demonstrates that the total \pt carried inside the jet cone is decreased in Pb+Pb events compared with that in \pp events~\cite{CMS-HIN-15-013}.
However, the modification of the jet's constituent particle \pt distributions, or any lower $\iptz$ selections, have not yet been studied.
 
This Letter presents a measurement of the yield of charged particles produced opposite in azimuth to a \Zboson boson with $\iptz > \SI{15}{\GeV}$ in \PbPb and \pp collisions at a nucleon--nucleon center-of-mass energy $\sqrt{s_\mathrm{NN}} = \SI{5.02}{\TeV}$ with the ATLAS detector at the LHC.
The \PbPb and \pp data were recorded in 2018 and 2017, respectively, and correspond to integrated luminosities of up to $\SI{1.7}{nb^{-1}}$ and $\SI{260}{pb^{-1}}$. The charged particles are required to have $\ptch > \SI{1}{\GeV}$ and be approximately back-to-back with the $Z$ boson in the transverse plane\footnote{ATLAS uses a right-handed coordinate system with its origin at the nominal interaction point (IP) in the center of the detector and the \(z\)-axis along the beam pipe. Cylindrical coordinates \((r,\phi)\) are used in the transverse plane, \(\phi\) being the azimuthal angle around the \(z\)-axis. The pseudorapidity is defined in terms of the polar angle \(\theta\) as \(\eta = -\ln \tan(\theta/2)\).}, with azimuthal separation $\Delta\phi$ larger than $3\pi/4$.
In simulations of \pp collisions, particles meeting these criteria reside primarily in the leading jet azimuthally opposite to the \Zboson boson.
The per-\Zboson yields of charged particles, $N_\mathrm{ch}$, are reported as a function of \ptch, $\left(1/N_Z\right) \left(d^2N_\mathrm{ch}/d\ptch d\Delta\phi\right)$, in \pp and \PbPb collisions.
To quantify the modification resulting from the partons' propagation through the QGP, the ratio of particle yields between \PbPb and \pp collisions, \IAA, is reported and compared with the expectations from theoretical calculations.
This measurement explores phenomena similar to those in measurements of the photon-tagged jet fragmentation function~\cite{HION-2017-08}.
However, requiring a reconstructed jet may result in a bias towards events with less energy loss than average~\cite{Agakishiev:2011nb,Renk:2012ve,Milhano:2015mng}.
Since there is no such requirement in this measurement, it provides additional insight into energy loss in an unbiased way, at low $p_\mathrm{T}^{Z/\gamma}$ values which have not yet been measured at the LHC and where theoretical models have not been tested.
 
\FloatBarrier
 
The ATLAS experiment~\cite{PERF-2007-01} is a multipurpose particle detector
with a forward--backward symmetric cylindrical geometry and a near \(4\pi\) coverage in
solid angle.
It consists of an inner tracking detector surrounded by a superconducting solenoid
providing a \SI{2}{\tesla} axial magnetic field, electromagnetic and hadron calorimeters, and a muon spectrometer.
The inner tracking detector covers the pseudorapidity range \(|\eta| < 2.5\).
It consists of silicon pixel, silicon microstrip, and transition radiation tracking detectors~\cite{ATLAS-TDR-2010-19-fixed,PIX-2018-001}.
Lead/liquid-argon (LAr) sampling calorimeters provide electromagnetic (EM) energy measurements
with high granularity.
A steel/scintillator-tile hadron calorimeter covers the central pseudorapidity range (\(|\eta| < 1.7\)).
Liquid-argon calorimeters with separate EM and hadronic compartments instrument the endcap (up to $\left|\eta\right| = 3.2$) and forward (FCal, up to $\left|\eta\right| = 4.9$) regions.
The muon spectrometer surrounds the calorimeters and includes
three air-core toroidal superconducting magnets with field integrals ranging between \num{2.0} and \SI{6.0}{\tesla\metre}, a system of precision tracking chambers, and fast detectors for triggering. During Pb+Pb data-taking, the muon system was operational for only 1.4~nb$^{-1}$ of the total integrated luminosity. Thus the dimuon channel is analyzed only in this subset of data.
 
\FloatBarrier
 
Events with a high-\pt electron or muon are initially selected for analysis by the single-lepton triggers described in Refs.~\cite{TRIG-2016-01,TRIG-2018-05}.
The centrality of \PbPb events is defined using the total transverse energy measured in the FCal~\cite{HION-2016-07,HION-2018-06}, \sumETPb.
\PbPb events are divided into three categories which correspond to the 0--10\%, 10--30\%, and 30--80\% centrality intervals in minimum-bias (MB) events, the smaller values indicating larger nuclear overlap regions and thus larger, hotter QGP regions.
The orientation of the underlying event (UE) elliptic flow is determined from the azimuthal distribution of the FCal energy~\cite{HION-2016-06,HION-2019-11-fixed}.
In \pp events, the average number of interactions per bunch crossing ranged from 2 to 4, and thus all charged-particle tracks are required to originate from the primary reconstructed vertex~\cite{ATL-PHYS-PUB-2015-018}.
 
Monte Carlo simulations of $\sqrt{s} = \SI{5.02}{\TeV}$ \pp collisions with $Z$ bosons decaying in the dielectron and dimuon channels, as well as data-driven studies, are used to correct the data for bin migration and reconstruction inefficiencies.
Generated events were passed through a \GEANT{4} simulation~\cite{Agostinelli:2002hh,SOFT-2010-01} of the ATLAS detector under the same conditions present during data-taking and were digitized and reconstructed in the same way as the data.
The $Z$ boson events were generated at next-to-leading order (NLO) with the \POWHEGBOXV{v2} program~\cite{Nason:2004rx,Frixione:2007vw,Alioli:2008gx,Alioli:2010xd} interfaced to the \PYTHIA 8.186 parton shower model~\cite{Sjostrand:2007gs}.
The NLO CT10 parton distribution function (PDF) set~\cite{Lai:2010vv} was used in the matrix element, while the CTEQ6L1 PDF set~\cite{Pumplin:2002vw} and the AZNLO tuned set of parameters~\cite{STDM-2012-23} were used to model the parton shower.
 
Four million events were generated to serve as the simulation sample for \pp collisions.
To model \PbPb events, fifteen million simulated \pp events were overlaid at the detector-hit level with MB \PbPb events in data.
This data-overlay sample was reweighted on an event-by-event basis to match the \sumETPb distribution for Pb+Pb events containing \Zboson bosons.
 
The \Zboson bosons in \pp and \PbPb events are reconstructed in opposite-sign dielectron and dimuon decay channels using procedures similar to those described in Refs.~\cite{HION-2018-10,HION-2018-02}. Reconstructed electrons are required to have a transverse momentum $\pte > \SI{20}{\GeV}$, to lie within the fiducial acceptance of the EM barrel ($\left|\eta^e\right| < 1.37$) or endcap ($1.52 < \left|\eta^e\right| < 2.47$) detectors, and to satisfy ``loose'' likelihood-based identification criteria, which have been optimized separately for \pp and \PbPb events~\cite{PERF-2017-01}.
Reconstructed muons are required to have a transverse momentum $\ptmu > \SI{20}{\GeV}$, to lie within the fiducial acceptance of the muon spectrometer ($\left|\eta^\mu\right| < 2.5$), and to pass the ``medium'' selection requirements described in Ref.~\cite{PERF-2015-10}.
The $Z \to \ell \ell$ candidates are required to be within the mass range $76 < m_{\ell \ell} < \SI{106}{\GeV}$ and have $\iptz > \SI{15}{\GeV}$.
This selection ensures that the contribution from multijet and other backgrounds is smaller than 1.5\% (0.1\%) for the dielectron (dimuon) channel, and is considered negligible.
In total, these criteria select approximately 21\,000 (28\,000) $Z\rightarrow{ee}$ ($Z\rightarrow\mu\mu$) events in \pp data, and 3400 (4100) events in Pb+Pb data.
 
Each \Zboson data event is assigned a series of weights, derived from simulation and data, to account for the trigger, reconstruction and selection efficiencies of its decay leptons.
Individual lepton trigger efficiencies are determined directly in \pp and \PbPb data using tag-and-probe techniques~\cite{TRIG-2018-05,TRIG-2016-01}, and are 0.70--0.80 for each muon and 0.75--0.95 for each electron.
Reconstruction and selection efficiencies are determined using simulation and are 0.65--0.80 for muons and 0.65--0.95 for electrons.
Although the efficiencies may vary substantially with the individual lepton \pt, $\eta$ and $\phi$, the resulting dependence on \iptz is weak due to the large $Z$ mass and weak correlation between bosons and their decay leptons.
 
Charged-particle tracks are reconstructed from hits in the inner detector using an algorithm~\cite{PERF-2015-08} which, in \PbPb collisions, is optimized for the high-occupancy conditions~\cite{HION-2011-03}.
They are required to meet several criteria intended to select primary charged particles~\cite{HION-2017-04}.
All reconstructed tracks with $\pt > \SI{1}{\GeV}$, $\left|\eta\right| < 2.5$ and $\Delta\phi > 3\pi/4$ are considered.
The charged-particle yield is corrected for reconstruction and selection inefficiency on a per-track basis using a simulation-derived efficiency which
varies from 0.6 to 0.8 depending on both detector occupancy and track kinematics.
A small correction, typically 1--2\%, accounts for the contribution of reconstructed tracks not associated with primary particles. The \ptch resolution is found to have a negligible effect ($\lesssim 0.3$\%) on the results and is not corrected for.
 
The contribution to the yield from UE particles in Pb+Pb collisions is estimated using MB events and is statistically subtracted from the measured yields.
For each \Zboson event in data, 40–-160 unique MB events are used for this estimation.
These MB events are centrality matched to within $1\%$ in peripheral events, decreasing to within $0.1\%$ in central events.
Furthermore, to match the azimuthal modulation of the UE, the elliptic flow angles~\cite{HION-2019-11-fixed} in the $Z$ data event and in the matching MB event must match within $\pi/16$.
The signal-to-background ratio varies strongly with \ptch, \iptz, and \PbPb centrality, with a minimum of $5\times10^{-3}$ at the lowest \ptch and \iptz values in the most central events. In \pp events, the UE is known to have larger activity in a \Zboson event than in an ordinary MB \pp collision \cite{CMS-FSQ-16-008,STDM-2017-28}, necessitating a different procedure. Here, the UE is determined in events with $1 < \iptz < 12$~\GeV\, in the azimuthal region perpendicular to the $Z$ boson to avoid the contribution from jet particles.
 
The data are further corrected for bin migration resulting from the finite resolution in the \iptz measurement.
This is evaluated by comparing the per-\Zboson charged-particle yields, where the \Zboson selection is made at the generator level, with those after reconstruction, and is typically a 2--3\% correction.
 
The primary sources of systematic uncertainty in the yield measurement are those affecting the \Zboson boson reconstruction, those affecting the charged-particle selection, and those affecting the UE background estimation and subtraction. The uncertainties associated with the electron and muon energy scales are evaluated using a common set of
uncertainties~\cite{PERF-2015-10,EGAM-2018-01}, and are typically negligible ($\lesssim 1$\%) except at high \ptch. Those associated with lepton trigger and selection efficiency determination are smaller than the ones related to the energy scale. Several sources of tracking-related uncertainty are considered, which are described in previous measurements of charged-particle fragmentation functions, and of which the largest is the sensitivity to the track selection criteria, which is $2$--$3$\%~\cite{HION-2017-04,PERF-2015-07}.
 
The uncertainty in the determination of the UE background yield is evaluated by propagating the statistical uncertainty of the UE estimation in MB events. The sensitivity of the UE estimation to the matching criteria for the elliptic flow~\cite{HION-2016-06} angles between signal and MB events, or the additional requirement to match the triangular flow angles, are investigated. However, since these variations give statistically compatible results, they are not included. As a check of the background subtraction procedure, the full analysis is performed on simulated $Z$ events overlaid with \textsc{Hijing}~\cite{Gyulassy:1994ew} Pb+Pb background, and compared with the generator-level distributions. An absolute uncertainty in the background estimation of $0.3$\% is derived using  this study.
 
Finally, an internal consistency check is performed by comparing the per-\Zboson yields between the electron and muon decay channels.
A difference was observed in the $15 < \iptz < \SI{30}{\GeV}$ selections and was included as an uncertainty of at most 4\% in \pp and 14\% in central \PbPb events.
 
For the yields at low \ptch and in central events, the uncertainty from the UE determination is dominant and can be as large as 30\%.
For yields at high \ptch and in lower-multiplicity events, the uncertainties associated with the track selection and the lepton energy scale are typically dominant, and as large as 5\%.
Uncertainty sources common to \PbPb and \pp are cancelled in the \IAA ratio when possible, such that the resulting measurement is dominated by uncertainties specific to \PbPb events.
In all cases, the statistical uncertainty in the \IAA is larger than the total systematic uncertainty.
 
\FloatBarrier
 
\begin{figure}[p]
\begin{center}
\centering
\includegraphics[width=0.50\columnwidth]{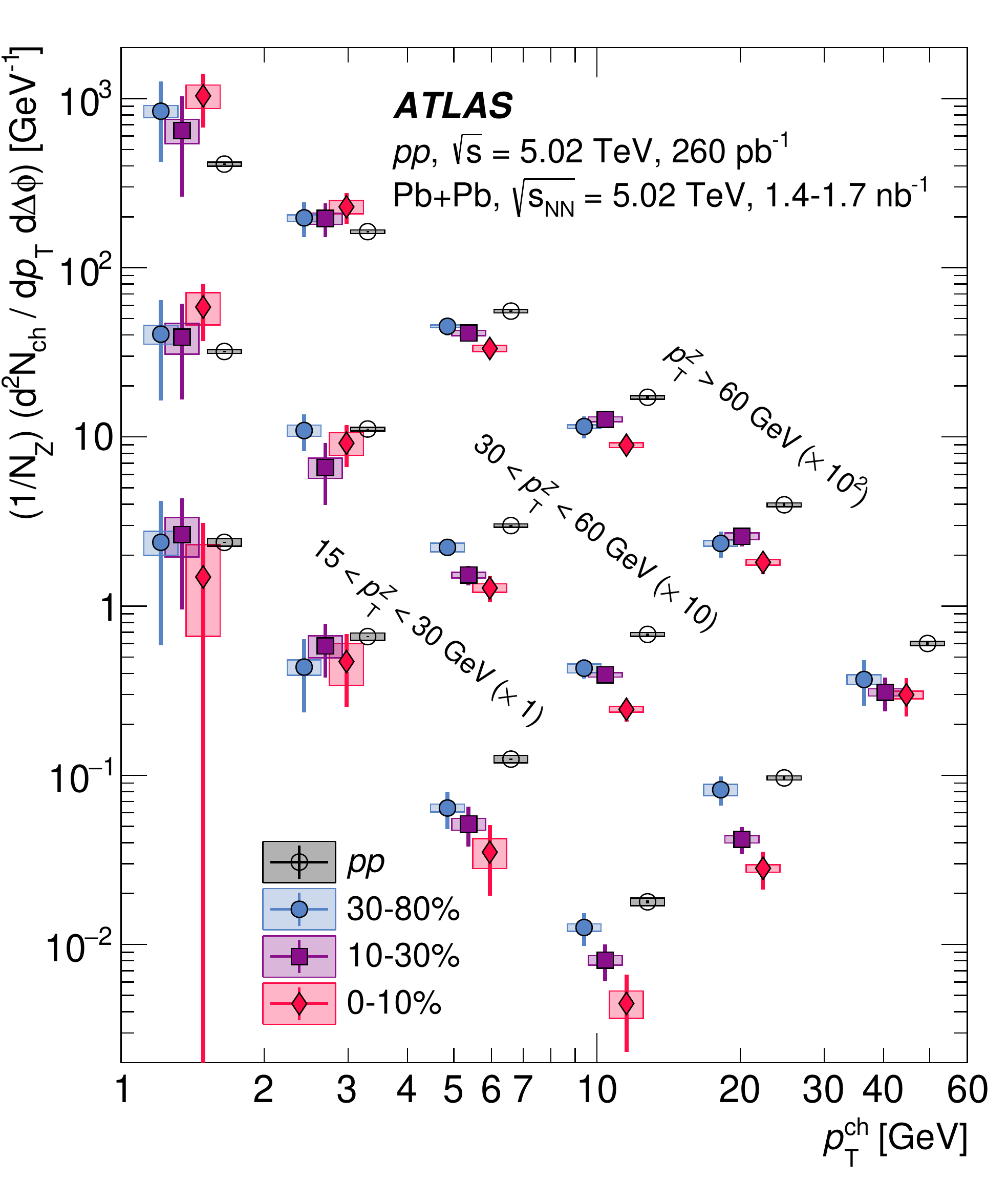}
\caption{Charged-particle yield per $Z$ boson as a function of \ptch, for the selection $\Delta\phi > 3\pi/4$, reported for $15 < \iptz < \SI{30}{\GeV}$, $30 < \iptz < \SI{60}{\GeV}$, and $\iptz > \SI{60}{\GeV}$. Results are shown for \pp events and the three centralities of \PbPb events. These are offset horizontally around the bin centers, which are located between the 0--10\% and 10--30\% points, for visibility. The vertical bars and boxes correspond to the statistical and systematic uncertainties of the data.
\label{fig1:yields_pTchOnly}}
\end{center}
\begin{center}
\centering
\includegraphics[width=0.50\columnwidth]{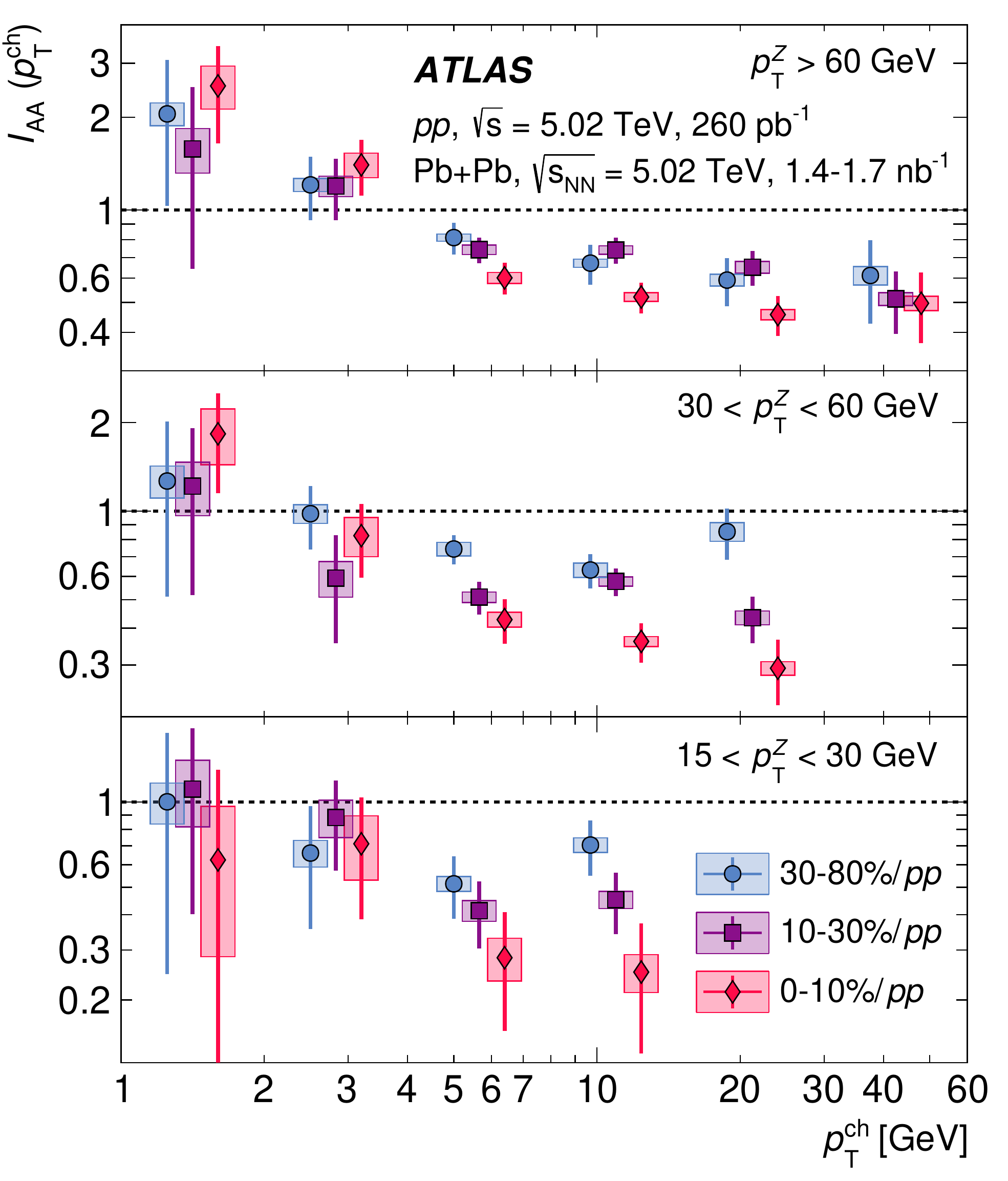}
\caption{Ratio of the charged-particle yield in \PbPb collisions to that in \pp collisions, \IAA, as a function of charged-particle \ptch, for the selection $\Delta\phi > 3\pi/4$. The vertical bars and boxes correspond to the statistical and systematic uncertainties of the data. The 0--10\% and 30--80\% data are offset horizontally for visibility.
\label{fig2:IAA_pTchOnly}}
\end{center}
\end{figure}

Figure~\ref{fig1:yields_pTchOnly} presents the charged-particle yield per $Z$ boson, in \PbPb and \pp events, as a function of \ptch, for the selection $\Delta\phi > 3\pi/4$. The yields in \PbPb collisions are observed to be modified relative to those in \pp collisions. To better reveal the modification, Figure~\ref{fig2:IAA_pTchOnly} presents \IAA values, the ratios of yields in \PbPb events to those in \pp events.
The \IAA values are suppressed below unity at large \ptch, with a systematically larger suppression in more central events and for lower \iptz selections.
For $\iptz > \SI{60}{\GeV}$, the $\IAA$ values at low $\ptch$, less than $2$--$\SI{3}{\GeV}$, are significantly different than those at high $\ptch$, and typically greater than unity.
Lower $\iptz$ selections are compatible with a similar increase at low $\ptch$, although the uncertainties limit the significance of this enhancement.
The suppression over a wide range of \ptch values, and the general enhancement of the \IAA above unity at lower \ptch, are qualitatively similar to those observed in the ratios of jet fragmentation functions in photon-tagged events~\cite{HION-2017-08}.

\begin{figure*}[!t]
\begin{center}
\centering
\includegraphics[width=0.99\linewidth]{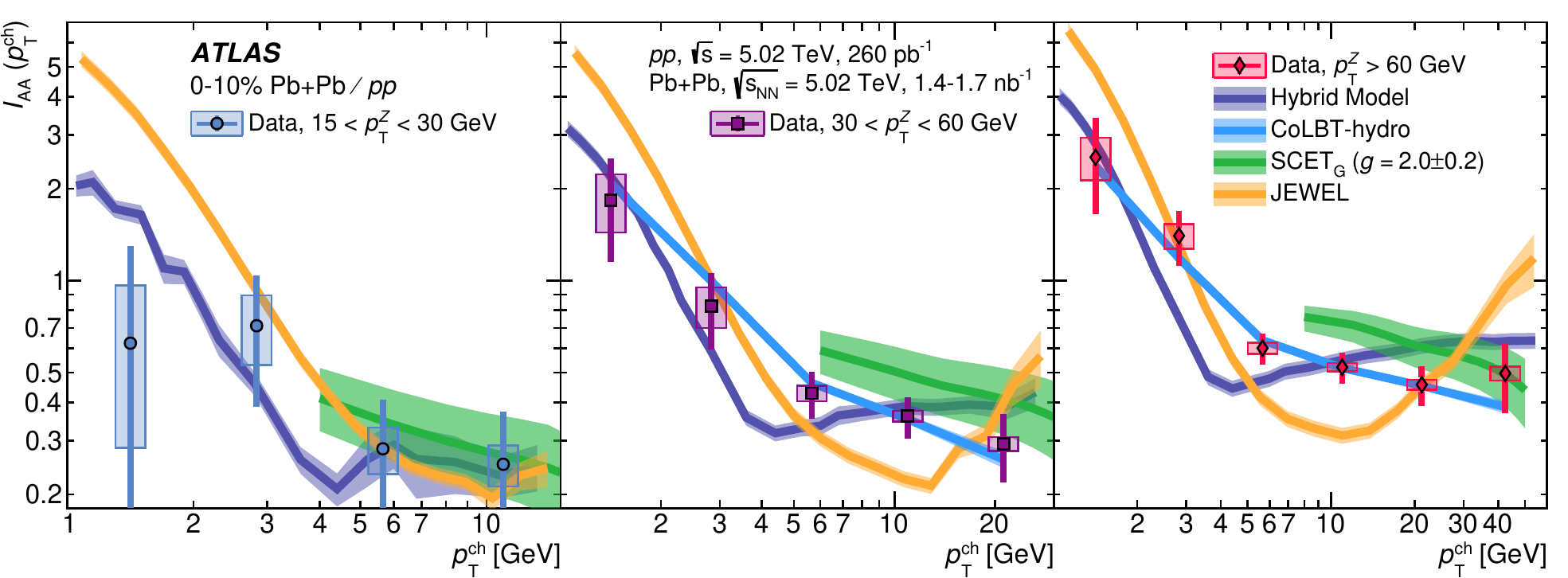}
\caption{The \IAA ratio as a function of \ptch in data compared with theoretical calculations (see text), for the selection $\Delta\phi > 3\pi/4$. The vertical bars and boxes correspond to the statistical and systematic uncertainties, while the shaded bands represent the theoretical uncertainty (statistical for JEWEL, Hybrid, and CoLBT-hydro, parametric for SCET$_\mathrm{G}$).
The \IAA is shown for 0--10\% Pb+Pb events for $\iptz = 15$--$30$~\GeV\ (left), $30$--$60$~\GeV\ (center) and $> 60$~\GeV\ (right).
\label{fig3:theory}}
\end{center}
\end{figure*}
 
Figure~\ref{fig3:theory} compares the \IAA in 0--10\% Pb+Pb events with the following theoretical calculations, where available, which use the same kinematic selections as the data:
(1) a perturbative calculation within the framework of soft-collinear effective field theory with Glauber gluons (SCET$_\mathrm{G}$) in the soft-gluon-emission (energy-loss) limit, with jet-medium coupling $g = 2.0 \pm 0.2$~\cite{Chien:2015vja,Li:2019dre}; (2) the Hybrid Strong/Weak Coupling model~\cite{Casalderrey-Solana:2015vaa}, which combines initial production using \PYTHIAV{8} with a parameterization of energy loss derived from holographic methods, including back-reaction effects;
(3) JEWEL, an MC event generator which simulates QCD jet evolution in heavy-ion collisions, including radiative and elastic energy loss processes, and configured to include medium recoils~\cite{KunnawalkamElayavalli:2016ttl}; and
(4) a coupled linearized Boltzmann transport (CoLBT) and hydrodynamics model~\cite{Chen:2017zte,Chen:2020tbl}, which  includes jet-induced medium excitations.
All models qualitatively reproduce the degree of suppression at large \ptch, greater than 10~\GeV.
The Hybrid model, JEWEL and CoLBT qualitatively capture the increase at low \ptch.
For these three models, removing the back-reaction, medium recoils, and jet-induced medium excitations, respectively, results in a significant underprediction of the data in this region.
Several of these models also capture the relative difference in the \IAA between the three \iptz selections.
A full evaluation of theoretical uncertainties is needed to further discriminate between the mechanisms of energy loss and medium response in the data.
 
\FloatBarrier
 
In conclusion, this Letter presents a measurement of charged-particle yields produced in the azimuthal direction opposite to a \Zboson boson with $\pt > \SI{15}{\GeV}$.
The measurement is performed using $\SI{260}{pb^{-1}}$ of \pp and up to  $\SI{1.7}{nb^{-1}}$ of \PbPb collision data at $\SI{5.02}{\TeV}$ with the ATLAS detector at the Large Hadron Collider.
The per-\Zboson yields are systematically modified in \PbPb collisions compared with \pp collisions due to the interactions between the parton shower and the hot and dense QGP medium.
The charged-particle \pt distribution in \PbPb collisions is softer than that in \pp collisions, with a suppression at high \ptch and an enhancement at low \ptch.
The degree of modification varies with \PbPb event centrality, consistent with a larger and hotter QGP being created in more central events.
At high \iptz, the modification pattern is qualitatively similar to that observed in measurements of photon-tagged jet fragmentation functions.
In addition to the particular theoretical comparisons presented here, the data will allow systematic tests of models across centrality and \iptz selections. The data can also test energy loss models for low-$p_\mathrm{T}$ partons that are otherwise difficult to access experimentally at the LHC, but which are valuable for direct comparison to future measurements at RHIC
 
\FloatBarrier
 
\section*{Acknowledgments}
We thank CERN for the very successful operation of the LHC, as well as the
support staff from our institutions without whom ATLAS could not be
operated efficiently.
 
We acknowledge the support of ANPCyT, Argentina; YerPhI, Armenia; ARC, Australia; BMWFW and FWF, Austria; ANAS, Azerbaijan; SSTC, Belarus; CNPq and FAPESP, Brazil; NSERC, NRC and CFI, Canada; CERN; CONICYT, Chile; CAS, MOST and NSFC, China; COLCIENCIAS, Colombia; MSMT CR, MPO CR and VSC CR, Czech Republic; DNRF and DNSRC, Denmark; IN2P3-CNRS and CEA-DRF/IRFU, France; SRNSFG, Georgia; BMBF, HGF and MPG, Germany; GSRT, Greece; RGC and Hong Kong SAR, China; ISF and Benoziyo Center, Israel; INFN, Italy; MEXT and JSPS, Japan; CNRST, Morocco; NWO, Netherlands; RCN, Norway; MNiSW and NCN, Poland; FCT, Portugal; MNE/IFA, Romania; MES of Russia and NRC KI, Russia Federation; JINR; MESTD, Serbia; MSSR, Slovakia; ARRS and MIZ\v{S}, Slovenia; DST/NRF, South Africa; MICINN, Spain; SRC and Wallenberg Foundation, Sweden; SERI, SNSF and Cantons of Bern and Geneva, Switzerland; MOST, Taiwan; TAEK, Turkey; STFC, United Kingdom; DOE and NSF, United States of America. In addition, individual groups and members have received support from BCKDF, CANARIE, Compute Canada and CRC, Canada; ERC, ERDF, Horizon 2020, Marie Sk{\l}odowska-Curie Actions and COST, European Union; Investissements d'Avenir Labex, Investissements d'Avenir Idex and ANR, France; DFG and AvH Foundation, Germany; Herakleitos, Thales and Aristeia programmes co-financed by EU-ESF and the Greek NSRF, Greece; BSF-NSF and GIF, Israel; La Caixa Banking Foundation, CERCA Programme Generalitat de Catalunya and PROMETEO and GenT Programmes Generalitat Valenciana, Spain; G\"{o}ran Gustafssons Stiftelse, Sweden; The Royal Society and Leverhulme Trust, United Kingdom.
 
The crucial computing support from all WLCG partners is acknowledged gratefully, in particular from CERN, the ATLAS Tier-1 facilities at TRIUMF (Canada), NDGF (Denmark, Norway, Sweden), CC-IN2P3 (France), KIT/GridKA (Germany), INFN-CNAF (Italy), NL-T1 (Netherlands), PIC (Spain), ASGC (Taiwan), RAL (UK) and BNL (USA), the Tier-2 facilities worldwide and large non-WLCG resource providers. Major contributors of computing resources are listed in Ref.~\cite{ATL-SOFT-PUB-2020-001}.
 
\FloatBarrier

\printbibliography
 
\clearpage
% ATLAS Collaboration author list
% Reference date of HION-2019-05 is 2020-06-22
% Author list last updated on date 23-FEB-21
% Data extracted on 23-Feb-2021 for paper reference HION-2019-05
% at 5:34pm
 
\begin{flushleft}
\hypersetup{urlcolor=black}
{\Large The ATLAS Collaboration}

\bigskip

\AtlasOrcid[0000-0002-6665-4934]{G.~Aad}$^\textrm{\scriptsize 102}$,    
\AtlasOrcid[0000-0002-5888-2734]{B.~Abbott}$^\textrm{\scriptsize 128}$,    
\AtlasOrcid[0000-0002-7248-3203]{D.C.~Abbott}$^\textrm{\scriptsize 103}$,    
\AtlasOrcid[0000-0002-2788-3822]{A.~Abed~Abud}$^\textrm{\scriptsize 36}$,    
\AtlasOrcid[0000-0002-1002-1652]{K.~Abeling}$^\textrm{\scriptsize 53}$,    
\AtlasOrcid[0000-0002-2987-4006]{D.K.~Abhayasinghe}$^\textrm{\scriptsize 94}$,    
\AtlasOrcid[0000-0002-8496-9294]{S.H.~Abidi}$^\textrm{\scriptsize 167}$,    
\AtlasOrcid[0000-0002-8279-9324]{O.S.~AbouZeid}$^\textrm{\scriptsize 40}$,    
\AtlasOrcid{N.L.~Abraham}$^\textrm{\scriptsize 156}$,    
\AtlasOrcid[0000-0001-5329-6640]{H.~Abramowicz}$^\textrm{\scriptsize 161}$
% ...
\bigskip
\\

% ...
$^{36}$CERN, Geneva; Switzerland.\\
% ...
$^{40}$Niels Bohr Institute, University of Copenhagen, Copenhagen; Denmark.\\
% ...
$^{53}$II. Physikalisches Institut, Georg-August-Universit\"{a}t G\"ottingen, G\"ottingen; Germany.\\
% ...
$^{94}$Department of Physics, Royal Holloway University of London, Egham; United Kingdom.\\
% ...
$^{102}$CPPM, Aix-Marseille Universit\'e, CNRS/IN2P3, Marseille; France.\\
$^{103}$Department of Physics, University of Massachusetts, Amherst MA; United States of America.\\
% ...
$^{128}$Homer L. Dodge Department of Physics and Astronomy, University of Oklahoma, Norman OK; United States of America.\\
% ...
$^{156}$Department of Physics and Astronomy, University of Sussex, Brighton; United Kingdom.\\
% ...
$^{161}$Raymond and Beverly Sackler School of Physics and Astronomy, Tel Aviv University, Tel Aviv; Israel.\\
% ...
$^{167}$Department of Physics, University of Toronto, Toronto ON; Canada.\\
% ...

\end{flushleft}

% Created with Glance <Atlas.Glance@cern.ch>

\end{document}